\documentclass[11pt,twocolumn]{article}

\usepackage{graphicx}
\usepackage{amsmath,amssymb,textcomp}
\usepackage{subfigure}
\usepackage{cite}
\usepackage[pdfstartview=FitH,colorlinks]{hyperref}
\usepackage{url,array}

\newcommand{\cm}{\text{cm}}
\newcommand{\pc}{\text{pc}}
\newcommand{\s}{\text{sec}}

\begin{document}

	\title{Gamma-ray effects of dark forces in dark matter clumps}

\author{K.~Belotsky$^{a,b}$\thanks{k-belotsky@yandex.ru},
\quad M.~Khlopov$^{a,b,c}$\thanks{khlopov@apc.univ-paris7.fr} and
\quad A.~Kirillov$^{a,b}$\thanks{kirillov-aa@yandex.ru} \\
$^a$ National Research Nuclear University ``MEPhI'', 115409, Moscow, Russia\\
$^b$ Centre for Cosmoparticle Physics ``Cosmion'', 115409, Moscow, Russia\\
$^c$ APC Laboratory, 10, rue Alice Domon et L\'{e}onie Duquet
75205 Paris Cedex 13, France}

\date{}

\twocolumn[
	\begin{@twocolumnfalse}
	\maketitle
	\abstract{
		Existence of new gauge U(1) symmetry possessed by dark matter (DM) particles implies the existence of a new Coulomb-like interaction, which leads to Sommerfeld-Gamow-Sakharov enhancement of dark matter annihilation at low relative velocities. We discuss a possibility to put constraints on the such dark forces of dark matter from the observational data on the gamma radiation in our Galaxy. Gamma-rays are supposed to originate from annihilation of DM particles in the small scale clumps, in which annihilation rate is supposed to be enhanced, besides higher density, due to smaller relative velocities $v$ of DM particles. For possible cross sections, mass of annihilating particles, masses of clumps and the contribution of annihilating particles in the total DM density we constrain the strength of new dark long range forces from comparison of predicted gamma ray signal with Fermi/LAT data on unidentified point-like gamma-ray sources (PGS) as well as on diffuse $\gamma$-radiation.
	}
	\bigskip


	\bigskip
  \end{@twocolumnfalse}
]
{
  \renewcommand{\thefootnote}%
    {\fnsymbol{footnote}}
  \footnotetext[1]{k-belotsky@yandex.ru}
  \footnotetext[2]{khlopov@apc.univ-paris7.fr}  
  \footnotetext[3]{kirillov-aa@yandex.ru}
}
\section{Introduction}

From the first articles revealing the indirect effects of the cold dark matter (CDM) in the form of heavy neutral leptons \cite{1977PhRvL..39..165L, 1977PhRvL..39..168D, 1978ApJ...221..327D, 1978ApJ...223.1015G, 1978ApJ...223.1032S, 1980YadPhys...31..1286, 1984PhRvL..53..624S, 1994PAN....57..425K} or supersymmetric particles \cite{1984PhRvL..53..624S, 1985PhRvL..55.2622S, 1988ApJ...325...16R}, such indirect effects of dark matter annihilation had been the subject of intensive studies in the data on the cosmic rays (CR) and gamma radiation. In the CDM scenario DM particles could form the hierarchic structures over a wide range of scales and masses (from small scale clumps to large scale structures) \cite{1977ApJ...218..592G, 1978MNRAS.183..341W, 1982ApJ...263L...1P, 1982Natur.299...37B, 1984Natur.311..517B, 1995SvPhU..38..687G}. The annihilation rate of DM particles within these clumpy structures, giving rise to cosmic ray signals \cite{1978ApJ...223.1015G, 1978ApJ...223.1032S, 1980YadPhys...31..1286, 1984PhRvL..53..624S, 1988PhLB..214..403E, 1991PhRvD..43.1774K, 1994PAN....57..425K, 1992PhLB..294..221B, 1995PhLA..208..276G, 2003PhRvD..68j3003B, 2005Natur.433..389D}, 
should be enhanced due to higher density of DM particles in them, as compared with their averaged density in the Galaxy. The annihilation cross section can be also enhanced at small relative velocities of DM particles, which are especially small in the lightest clumps, which are likely to be the most abundant. Due to these factors the clumps, located in a neighborhood of Solar system, can be observed as discrete (basically point-like) gamma-ray sources \cite{1992PhLB..294..221B, 1997SvPhU..40..869G, 2003PhRvD..68j3003B, 2013PAN....76..469B, 2012arXiv1212.6087B}, while the overall effect of such clumps should increase the diffuse gamma background in the Galaxy.

In this paper we address the sensitivity of the given effect \cite{2013PAN....76..469B, 2012arXiv1212.6087B} to the existence of long range dark forces originated from new U(1) gauge charges, which annihilating dark matter particles possess. Coulomb-like interaction (we will refer to it as ``y-interaction'') of these U(1) charges leads to Sommerfeld-Gamow-Sakharov (SGS) enhancement of particle annihilation at low relative velocities. Since small scale clumps of dark matter have small gravitational potential the relative velocities of particles, annihilating within these clumps, is much smaller, than of unclumped particles in the galactic halo, what enhances the sensitivity of both the data on the discrete gamma-ray sources and on the diffuse $\gamma$-radiation to the effects of the dark forces in dark matter annihilation. 

One should note that effects of DM annihilation enhancement due to new interaction were first considered in \cite{2000GrCo....6..140, 2005GrCo...11...27B} and then in many successive papers, e.g. \cite{2012JCAP...01..041A, 2012PhRvD..86h3534S, 2012PhRvD..85l3512V, 2012JCAP...12..009A, 2013JCAP...04..017C}. In most of these papers SGS enhancement was mainly studied for the case of dominant form of DM that possess new interaction with massive carrier, as well as the effect of this enhancement for point-like gamma-ray sources, treated in the present work,  were not considered. 

The existence of new U(1) gauge symmetry implies the existence of the corresponding massless dark photons. They can increase the effective number of relativistic species at the radiation dominated stage. Decoupling of such photons from the plasma in the early Universe takes place in the period of freezing out of the CDM particles, what makes this contribution compatible with the measurements of the effective number of the neutrino species (see e.g. \cite{2006GrCo...12...93B} for review and references).

Note that new interaction can lead to binding of DM particles into atomic-like states with their successive annihilation \cite{2006GrCo...12...93B}, or stable dark atoms in case of multicomponent (charge asymmetric) DM \cite{2006GrCo...12...93B, 2010JCAP...05..021K, 2013PhRvD..87j3515C}. This effect, going beyond the scope of the present work, deserves separate consideration.

\section{Gamma-ray signal from y-interacting DM clumps}
\label{sec:flux}

In calculations of gamma-ray signals from the clump in this paper we follow our previous work \cite{2013PAN....76..469B, 2012arXiv1212.6087B}.
For density profile inside the clumps 
we use profile BGZ obtained in \cite{1992PhLB..294..221B, 1997SvPhU..40..869G, 2003PhRvD..68j3003B}, which gives the minimal estimation of the $\gamma$-flux from the clump \cite{2013PAN....76..469B, 2012arXiv1212.6087B}. 
%
Fraction $\xi = 0.002$ from total density of DM in Galaxy is taken for the clumps survived until present time. We study $\gamma$-radiation effect for only minimal clump mass, formally assuming that all DM clumps are with this mass. 
According to theoretical estimates consistent with observations, the small mass clumps are predicted to be the most abundant \cite{1992PhLB..294..221B, 2003PhRvD..68j3003B, 2010ApJ...723L.195I}. However, the following merging and formation of the high-mass clumps (sub-halos) lead to the transformation of the mass distribution and the sub-halos effect can be also noticeable \cite{2012ApJ...747..121A}, but the picture does not change a lot.


To cover a wide class of models of DM particles, we parametrize their annihilation cross section as follows:
\begin{equation}
	\sigma_{\text{ann}}=\frac{\sigma_0}{v}\times C(v,\alpha),
	\label{sigma_common}
\end{equation}
what corresponds to the $s$-wave amplitude only. Parameter $\sigma_0$ is determined by cosmological density of the particles $\Omega$. The factor $C(v,\alpha)$ takes explicitly into account a possible Coulomb-like y-interaction of DM particles, which leads to a Sommerfeld-Gamow-Sakharov enhancement \cite{Sommerfeld, 1928ZPhy...51..204G, 1931AnP...403..257S, 1948ZETF...18..631} and has the form:

\begin{equation}
    C\left(v,\alpha\right) = \frac{2\pi\alpha/v}{1-\exp\left(-2\pi\alpha/v\right)}.
    \label{eq:coulomb}
\end{equation}
Here $\alpha$ is the fine structure constant of additional interaction.

One should note that y-interaction implies not only SGS enhancement of the annihilation cross section but also leads to the existence of new channels that involve y-photons in the final state. The corresponding suppression of the branching ratio for ordinary photon production is effectively taken into account in our calculations by a multiplicity of produced photons, $N_\gamma$ (see below).

The enhancement (eq.~\eqref{eq:coulomb}) of the annihilation cross section may lead to decrease of the frozen out density of DM particles. However in the period of their freezing out in the early Universe the particles were semirelativistic (with typical velocities $v \sim c/5$) so that relic density cannot decrease significantly. Effect of annihilation of DM particles in massless bosons of $y$-interaction ($y$ photons) needs in general special study in the framework of particular models of DM particles but it also cannot strongly increase the annihilation cross section (and correspondingly decrease the relic abundance) taking into account all the other possible annihilation channels.

In the modern Universe, when the particle velocities are nonrelativistic, the factor (eq.~\eqref{eq:coulomb}) may significantly enhance the annihilation effects  \cite{2010PhLB..687..275D, 2010PhRvD..81h3502Z, 2010PhRvD..82h3525F, 2011PhRvD..83l3513Z}. Such effects become noticeable even for a subdominant component with $\Omega\ll\Omega_{\text{CDM}}$ as it takes place in case of heavy stable neutrinos with $y$ interaction \cite{2008PAN....71..147B}.
Therefore we suppose that an active (annihilating) component of DM may be both dominant and subdominant, i.e. $\Omega\le\Omega_{\text{CDM}}$ with $\Omega_{\text{CDM}}\sim 0.2$ being the total relative density of cold dark matter in Universe. In estimation of cosmological density $\Omega$ we follow to the standard approach \cite{1981RvMP...53....1D, 1986PhRvD..33.1585S}.

It is worth to note that the given scheme does not take into account possibility of binding pairs of considered particle-antiparticle due to $y$-interaction. If nonrelativistic DM particles are decoupled from the ambient plasma their rapid cooling can strongly enhance the rate of such recombination that may exceed the expansion rate due to high recombination cross section. This process leads inevitably to annihilation and it may strongly suppress the abundance of these DM particles \cite{2005GrCo...11...27B}.


Since the Coulomb-like $y$-interaction is excluded for Majorana particles, it is assumed in our estimations that the considered DM particles are Dirac particles with mass $m\sim 100$~GeV. If annihilating DM particles do not constitute all DM ($\Omega<\Omega_\text{CDM}$) then their contribution to density of clumps is assumed to be proportional to $\Omega/\Omega_\text{CDM}$ (eq.~\eqref{eq:concentration}). We do not specify annihilation channel of photon production, assuming that their averaged multiplicity for energy $E_{\gamma}>100$~MeV is $N_\gamma = 10$. It is quite typical value for high energy processes at respective energy release.


The photon flux at distance $l$ from the clump centre is given by
\begin{equation}
	F = \frac{P}{4\pi l^2}= \frac{N_\gamma}{4\pi l^2} \int\limits_{V} \left\langle \sigma_\text{ann}v \right\rangle n\bar{n}\ dV,
	\label{eq:flux}
\end{equation}
where the particles/antiparticles number density is
\begin{equation}
    n = \bar{n} = \frac{1}{2} \frac{\rho(r)}{m}\times \frac{\Omega}{\Omega_{\text{CDM}}}.
    \label{eq:concentration}
\end{equation}
Note that the fraction of subdominant DM particles should be suppressed in the clumps of mass $M<M_{\rm min}$, if they are, where $M_{\rm min}$ is the minimal mass which could be formed by considered DM particles if they prevailed in density. In our study we do not take into account this.
The value $\left\langle\sigma_\text{ann}v\right\rangle$ is determined by averaging over velocity distribution of DM particles inside the clump, assumed to be Maxwellian one with the ``virial'' temperature $T_\text{vir}=GMm/2R$.

LAT registers $\gamma$-radiation with energy $E_{\gamma}>100$~MeV \cite{2009ApJ...697.1071A} and the flux 
$F>F_{\text{min}}\approx 3\times 10^{-9}$~$\cm^{-2}\s^{-1}$. The value $F_{\text{min}}$ determines the maximal distance $l_\text{max}$ at which the clump can be registered as $\gamma$-source. It gives for chosen profile $l_\text{max}\sim 10^{-5}$~pc for $\sigma_0=10^{-35}$~cm$^2$ and $10^{-10}M_\odot$ without y-interaction and $l_\text{max} \sim 1$~pc with one and the same parameters. All the obtained $l_\text{max}\ll$~Galactic size, what justifies assumption that clump number density $n_\text{cl}\approx\text{const}$ and corresponds to the local one. So
\begin{equation}
	n_{\text{cl}}=\frac{\xi \rho_{\text{loc}}}{M} \approx 1.6\times 10^{-5} \frac{M_\odot}{M}\ \pc^{-3}
	\label{eq:n_cl}
\end{equation}
where $\rho_\text{loc}=0.3$~GeV/cm$^3$.
The number of clumps which may be detected by LAT is
\begin{equation}
    N_{\text{cl}} =
    n_{\text{cl}}\times \frac{4}{3} \pi l_{\text{max}}^3.
    \label{eq:N}
\end{equation}

The analogous results for some other profile models are given in \cite{2012arXiv1212.6087B}. 

\begin{figure}[t]
  \centering
  \includegraphics[width=0.48\textwidth]{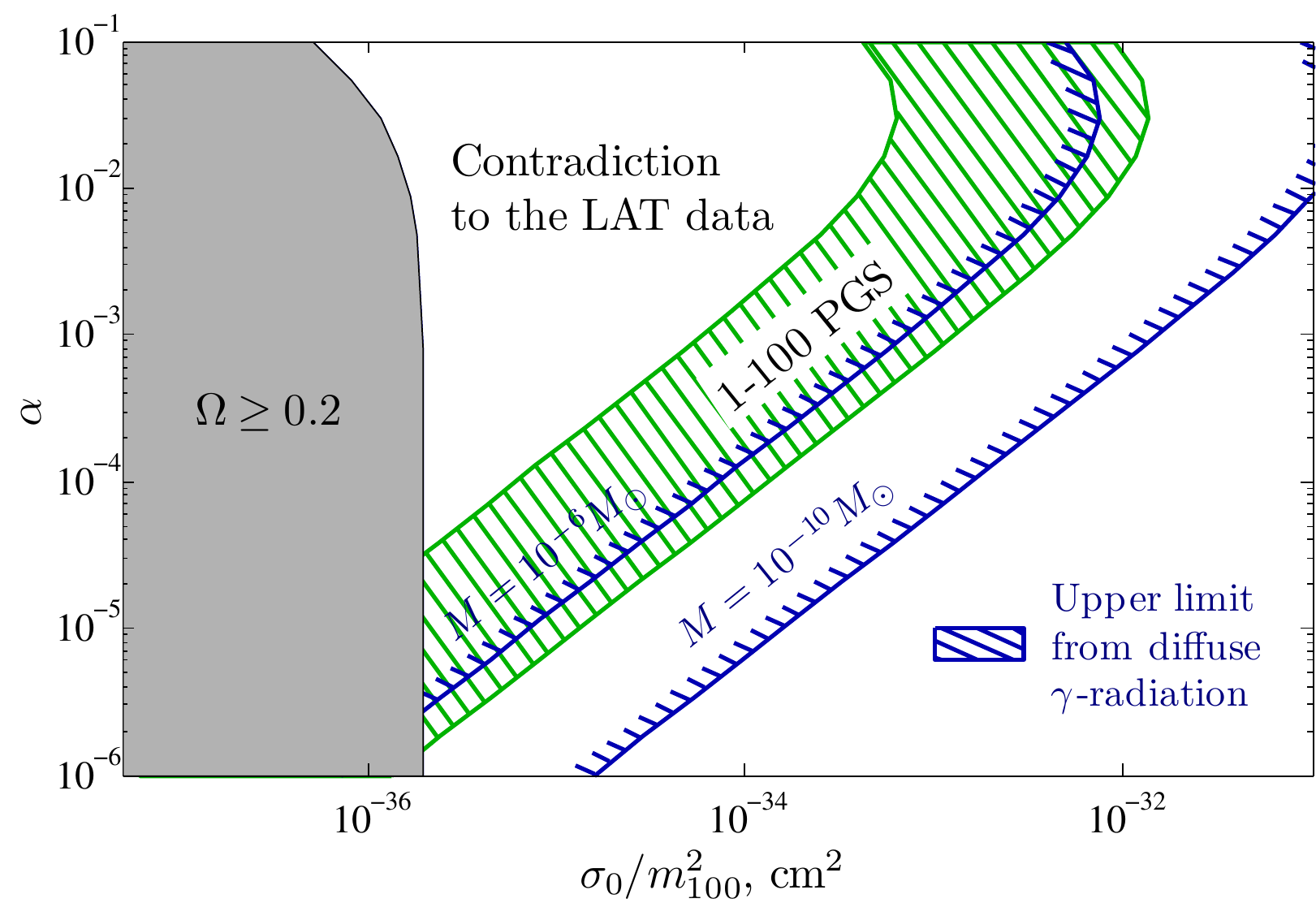}
  \caption{Allowed and forbidden regions of parameters $\alpha$ and $\sigma_0$ are shown as they obtained on the base of Fermi LAT data on PGS and diffuse $\gamma$-radiation. Here $m_{100}$ is the dark mater particle mass $m$ in units of 100~GeV.}
  \label{fig:alpha}
\end{figure}

Since the clumps at distance $l<l_\text{max}$ are expected to be distributed homogeneously, we can try to explain by them only isotropic component of unidentified PGS registered by Fermi LAT \cite{2012ApJS..199...31N}. It includes $\sim$100 sources. From this, 
respective regions of the parameters $\alpha$ and $\sigma_0$ for the typical clump masses $10^{-10} \div 10^{-6} M_{\odot}$ \cite{2005Natur.433..389D} are obtained. The results are shown at fig.~\ref{fig:alpha}. Note, that unidentified PGS data also allow another application to the exotic physics \cite{2011GrCo...17...27B, 2011APh....35...28B}.

The distant clumps situated at $l> l_\text{max}$ should contribute in the diffuse $\gamma$-radiation. $\gamma$-Flux from them per given solid angle can be expressed as
\begin{equation}
	\Phi= \int\limits^{l_{\rm halo}}_{l_\text{max}} F n_{\text{cl}} l^2 dl = \frac{P n_{\text{cl}}l_{\rm halo}^{\rm eff}}{4\pi},
	\label{eq:Phi}
\end{equation}
where $F$ and $P$ are introduced in Eq.~\ref{eq:flux}, $l_{\rm halo}$ is the distance to the edge of halo along to line of sight and $l_{\rm halo}^{\rm eff}\approx 10$~kps is its effective value (typical for many halo density profiles); $l_\text{max}$ is negligible with respect to $l_{\rm halo}$.
One requires that
\begin{equation}
    \Phi<\Phi_{\rm exp}\approx 1.5\times 10^{-5}\ \text{cm}^{-2}\text{s}^{-1}\text{sr}^{-1},
    \label{eq:Phi_exp}
\end{equation}
where $\Phi_{\rm exp}$ is the diffuse $\gamma$-background measured by LAT \cite{2009PhRvL.103y1101A}. 
It puts the upper limits on the annihilation cross section parameters which are also plotted on fig.~\ref{fig:alpha}.
It follows from these constraints that the case of clump mass $M=10^{-10}M_{\odot}$ is completely ruled out, while the case of $M=10^{-6}M_{\odot}$ is constrained but up to $\sim$10 PGS are still possible. Higher masses of clumps avoid these restrictions.


As to the parameters $\sigma_0$ and $\alpha$, the range $3\times 10^{-36}\lesssim\sigma_0\lesssim 10^{-32}$~cm$^2$ and $10^{-6}\lesssim\alpha\lesssim 10^{-1}$ 
are found to be the most interesting (fig.~\ref{fig:alpha}).

Effects of DM annihilation during period of the recombination of hydrogen can put constraints on the parameters of $y$ interaction. However, this question should be considered along with effects of DM recombination which might help to escape these constraints, what is out of the scope of the present paper.

\begin{figure}[t]%
	\centering
	\includegraphics[width=0.5\textwidth]{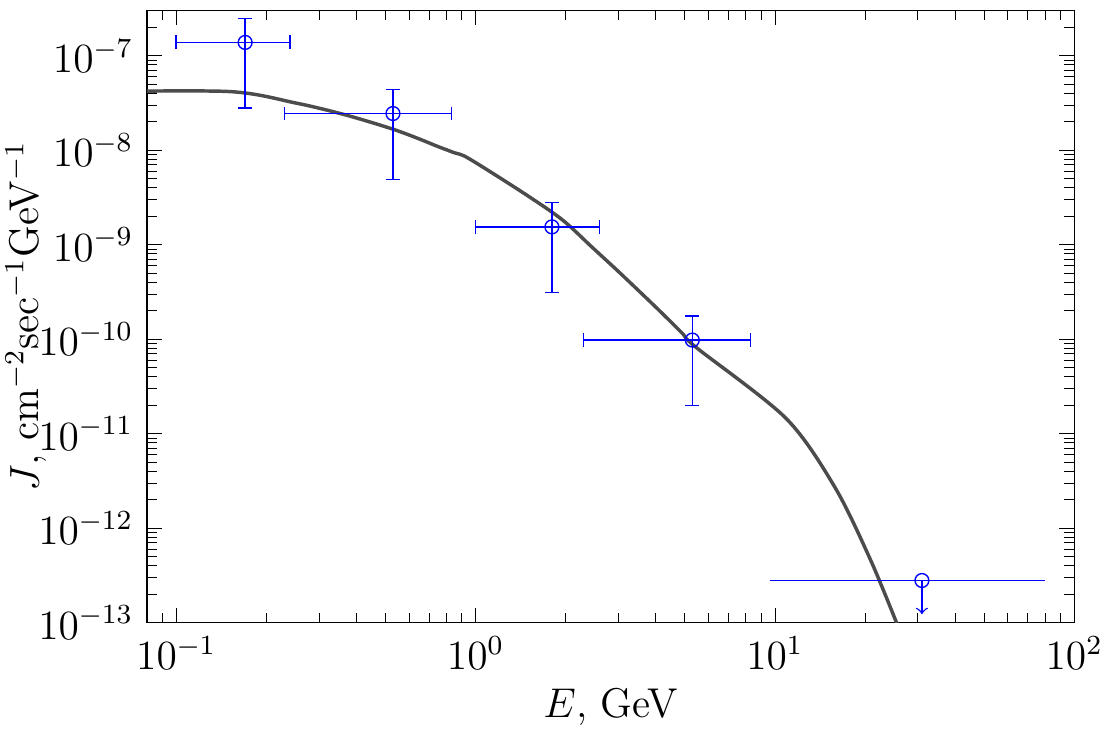}%
	\caption{Expected spectrum from $\nu_4\bar\nu_4$-annihilation
    is shown for $m_{\nu_4}=47$~GeV in comparison with the observed spectrum of unidentified source 2FGL J1653.6-0159.}%
	\label{fig:NeutrinosSpectrum}%
\end{figure}

It is interesting to note that one of the subdominant dark matter candidates~--- heavy neutrinos $\nu_4$ with y-interaction -  can explain a part of unidentified point-like LAT sources  due to $\nu_4\bar{\nu}_4$-annihilation with mass $m_{\nu_4}\sim 46-49$~GeV \cite{2012arXiv1212.6087B, 2013PAN....76..469B}. At the fig.~\ref{fig:NeutrinosSpectrum} a typical $\gamma$-spectrum from 47~GeV neutrinos annihilation is shown in comparison with measured spectrum of one of the non-identified PGS (annihilation spectrum was obtained with the help of Monte-Carlo generator Pythia 6.4 \cite{url:Pythia}). However, one should take into account that the heavy neutrino parameters are strongly restricted by underground experiments \cite{2008PAN....71..147B, 2009PhRvL.102a1301A}, as well as the predicted $\nu_4$ relic density suffers with an uncertainty related with their possible annihilation due to recombination of the y-interacting neutrinos and antineutrinos after their freezing out in the early Universe \cite{2008PAN....71..147B}.

\section{Conclusion}

In this paper we apply the approach developed earlier in our analysis of $\gamma$-radiation from annihilation of dark matter in clumps \cite{2013PAN....76..469B, 2012arXiv1212.6087B} to the case of new U(1) symmetry and put constraints on the parameters of the corresponding dark force from the data on discrete gamma sources and gamma background.  We have shown that DM clumps in vicinity of the Solar system could be observed as point-like sources of the $\gamma$-radiation and they can partially explain unidentified $\gamma$-sources, registered by LAT. 
Effects of DM annihilation in more distant clumps contribute to the diffuse gamma background and this contribution strongly depends on the minimal mass of the clumps. The smaller is the mass of clumps the stronger is effect of Sommerfeld-Gamov-Sakharov enhancement of the annihilation rate so that the observational data on $\gamma$-background and on the unidentified point-like $\gamma$-sources provide constraints on the strength of the Coulomb-like dark force. 

These constraints are highly sensitive the choice of a density profile inside the clump and are obtained for the most ``conservative'' BGZ \cite{2003PhRvD..68j3003B} model.
It should be also noted that the suppression of the subdominant fraction of DM particles in clumps of mass $M<M_\text{min}$ has not been taken into account and requires special study.




\section{Acknowledgements}

The authors express gratitude to V.~Dokuchaev and Yu.~Eroshenko for useful discussion of clump evolution. The work of A.K. and K.B. was supported by grant of RFBR \textnumero~12-02-12123 and by The Ministry of education and science of Russia, projects \textnumero~14.132.21.1446, \textnumero~8525 and \textnumero~14.A18.21.0789.

\bibliographystyle{unsrt}
\bibliography{Article}

\end{document}